\documentclass[prd,aps,twocolumn,nofootinbib,superscriptaddress,eqsecnum,floatfix,preprintnumbers,amsmath,amssymb,longbibliography]{revtex4-2}

\makeatletter
\renewcommand{\p@subsection}{}
\renewcommand{\p@subsubsection}{}
\makeatother

\usepackage{graphicx}
\usepackage{epsfig}
\usepackage[dvipsnames]{xcolor}
\usepackage[utf8]{inputenc}
\usepackage{stmaryrd}
\usepackage{mathrsfs}
\usepackage{mathalfa}
\usepackage{accents}
\usepackage{enumitem}
\usepackage[normalem]{ulem}

\usepackage[colorlinks=true,
            citecolor=green,
            linkcolor=red,
            filecolor=cyan,
            urlcolor=magenta,
            backref=false]{hyperref}

\newcommand{\be}{\begin{equation}}
\newcommand{\ee}{\end{equation}}
\newcommand{\bea}{\begin{eqnarray}}
\newcommand{\eea}{\end{eqnarray}}
\newcommand{\ba}{\begin{equation}\begin{aligned}}
\newcommand{\ea}{\end{aligned}\end{equation}}
\newcommand{\beg}{\begin{gather*}}
\newcommand{\eng}{\end{gather*}}
\newcommand{\hh}{,\hspace{0.5cm}}
\newcommand{\hhh}{,\hspace{0.2cm}}

\newcommand{\n}[1]{\label{#1}}
\newcommand{\ins}[1]{{\mbox{\tiny #1}}}

\newcommand{\MC}[1]{{\mathcal{#1}}}
\newcommand{\CAL}{\mathcal}

\newcommand{\bs}{\begin{split}}
\newcommand{\es}{\end{split}}

\begin{document}

\title{Vaidya-Type Solutions of Quasitopological Gravity Interacting with Nonlinear Electrodynamics}

\author{Valeri P. Frolov}%
\email[]{vfrolov@ualberta.ca}
\affiliation{Theoretical Physics Institute, Department of Physics,
University of Alberta,\\
Edmonton, Alberta, T6G 2E1, Canada
}

\author{Chul-Moon Yoo}%
\email[]{yoo.chulmoon.k6@f.mail.nagoya-u.ac.jp}
\affiliation{Graduate School of Science, Nagoya University,
Nagoya 464-8602, Japan}
\affiliation{Kobayashi-Maskawa Institute for the Origin of Particles and the Universe (KMI),\\
Nagoya 464-8602, Japan
}

\author{Andrei Zelnikov}%
\email[]{zelnikov@ualberta.ca}
\affiliation{Theoretical Physics Institute, Department of Physics,
University of Alberta,\\
Edmonton, Alberta, T6G 2E1, Canada
}


\begin{abstract}
We construct Vaidya-type solutions of quasitopological gravity coupled to nonlinear electrodynamics in arbitrary spacetime dimensions. Starting from the corresponding static spherically symmetric charged solutions, we obtain their dynamical counterparts by promoting the integration constants, in particular the mass and electric charge, to arbitrary functions of the advanced or retarded null coordinate. We show that the resulting field equations are satisfied provided suitable charge-carrying null currents and null fluid fluxes are included. The formalism applies to a broad class of nonlinear electromagnetic theories and provides a simple and systematic method for generating exact radiating charged solutions in quasitopological gravity.

\hfill    Alberta Thy 1-26

\hfill       NU-QG-23
\end{abstract}

\maketitle

\section{Introduction}

In this paper, we investigate Vaidya-type solutions of quasitopological gravity (QTG) coupled to nonlinear electrodynamics. We show that these time-dependent solutions arise naturally by promoting the integration constants of the corresponding static configurations, in particular the mass and electric charge, to functions of the advanced or retarded null coordinate. This construction is a straightforward generalization of the procedure originally used to obtain the Vaidya solutions in Einstein gravity.

Our approach closely parallels the construction of the classical Vaidya solution in Einstein gravity. The key observation is that the field equations remain consistent when the parameters characterizing the corresponding static geometry are promoted to arbitrary functions of a null coordinate. The resulting time dependence is supported by physically natural matter sources describing null radiation together with electric currents carrying the varying electric charge. We derive the corresponding stress-energy tensor and electric current for a general nonlinear electromagnetic theory in arbitrary spacetime dimension and show that the total stress-energy tensor is conserved.

The formalism is illustrated using the recently obtained static charged black-hole solutions of quasitopological gravity coupled to nonlinear electrodynamics \cite{PinedoSoto:2026hfm}. This yields a broad family of exact radiating charged solutions describing black holes emitting or absorbing both neutral and charged null radiation. These solutions provide a useful framework for studying dynamical processes, including gravitational collapse, accretion, and black-hole evaporation, in higher-curvature gravity.

The recently proposed and actively studied theory of QTG represents a particular class of modified gravity models \cite{Bueno:2020gq,Bueno:2019ltp,Bueno:2022res,Oliva_2010,Hennigar:2017ego,Myers:2010ru,Moreno:2023rfl,QT_BH,rbh_pfkz,PinedoSoto:2025hel, Bueno:2025tli, Bueno:2025zaj, Frolov:2025ddw, Bueno:2026dln,Sueto:2026epz,Borissova:2026krh,Borissova:2026wmn,Borissova:2026rbi,Bueno:2025qjk}. It is constructed by supplementing the Einstein--Hilbert action with higher-order in curvature terms
\begin{equation}
\label{QTA}
\begin{split}
&S_\ins{QTG}=\frac{1}{2\varkappa}\,\int d^D x \,\sqrt{-g}\,L_\ins{QTG}\, ,\\
&L_\ins{QTG}= R + \sum_j \alpha_j \ell^{2(j-1)} Z_j \, .
\end{split}
\end{equation}
Here $\varkappa=8\pi G^{D}$, where $G^D$ is the $D$-dimensional gravitational coupling constant.
The quantities $Z_j$ entering the action are specially constructed scalar invariants, each given by a polynomial of order $j$ in the curvature. It has been shown that these invariants can be chosen such that, when the action is restricted to spherically symmetric spacetimes, the resulting field equations contain no derivatives higher than second order.

The explicit form of the curvature polynomials $\CAL{Z}_j$, together with the recursive relations defining their construction, can be found in \cite{Bueno:2020gq}.
The parameter $\ell$, which has dimensions of length, sets the fundamental scale at which higher-curvature corrections become significant.
The dimensionless coefficients $\alpha_j$ specify a particular model within this class of theories.
It was further demonstrated that, if the series in~\eqref{QTA} is not truncated and the coefficients $\alpha_j$ satisfy appropriate conditions, the corresponding spherically symmetric solutions in QTG describe regular black holes.

The study of regular black holes has recently become a ``hot" topic, attracting considerable attention and leading to a large number of publications. The reason is straightforward: general relativity predicts the existence of spacetime singularities inside black holes, a feature that is widely regarded as an indication that this theory is incomplete.

A number of phenomenological regular black-hole metrics have been proposed by replacing the Schwarzschild geometry in the high-curvature region with a nonsingular core while preserving the asymptotic behavior of the spacetime. Representative examples include \cite{Bardeen1968,FrVi,Dymnikova1992,Hayward2006,Frolov2016}. These metrics were originally introduced as geometrical ansatz rather than as solutions derived from an underlying gravitational action. More information and related references can be found in the following review articles \cite{Lan2023RBHreview, Bambi2023,nonsingularparadigmblackhole}.
Such models possess an intriguing feature: the black hole interior may contain one or more nonsingular regions that can be interpreted as new universes or expanding cosmological domains \cite{FMM_1, FMM_2,FrBa,DymnikovaDoboszFilchenkovGromov2001}. Charged black holes in QTG were recently discussed in \cite{hao2026chargedregularblackholes,Carballo-Rubio:2026mvj}. Regular black hole solutions have also been studied in the context of two-dimensional dilaton gravity. (see e,g \cite{collieaux2018,PhysRevD:lemos,PhysRevD:Kunstatter, PhysRevResearch:frolov}).

One of the main attractions of quasitopological gravity is that regular black hole solutions arise naturally from its field equations, without the need to modify the metric by hand \cite{QT_BH,bueno2025nonpolynomial, Myers:2010ru, rbh_pfkz, Bueno:2025tli, Bueno:2025zaj,PinedoSoto:2025hel,Bueno:2026dln}. In this sense, QTG provides a self-consistent framework in which the resolution of black hole singularities emerges as a consequence of the underlying gravitational dynamics rather than as an externally imposed assumption.
These models may offer a new perspective on the mass inflation problem by providing a self-consistent framework in which the internal structure of regular black holes can be investigated
\cite{Frolov:2026rcm,DiFilippo:2026jpv,DiFilippo:2024mwm}.

In this paper we consider quasitopological gravity (QTG) coupled to an electromagnetic field described by nonlinear electrodynamics. Our starting point is the recently obtained family of static, spherically symmetric solutions of this theory \cite{PinedoSoto:2026hfm}. These solutions are characterized by two integration constants, interpreted as the mass and electric charge of the black hole. We show that, after expressing the metric in advanced or retarded null coordinates and promoting these integration constants to functions of the corresponding null coordinate, one obtains a natural Vaidya-type generalization of the static solutions. The resulting geometry is sourced by a charge-carrying null current and a null energy flux, which are responsible for the evolution of the black hole charge and mass. This construction provides a simple and systematic method for generating dynamical charged solutions in QTG coupled to nonlinear electrodynamics.

The paper is organized as follows. In Section\,\ref{Sec2}, we derive the field equations for a general class of spherically symmetric Vaidya-type spacetimes and obtain the corresponding conservation laws in the presence of null currents and energy fluxes. In Section\,\ref{Sec3}, we establish a generalized Birkhoff theorem in quasitopological gravity for matter sources whose stress-energy tensor satisfies a particular set of conditions.

In Section\,\ref{Sec4}, we apply the general formalism to quasitopological gravity coupled to nonlinear electrodynamics and construct the corresponding charged Vaidya-type solutions by promoting the mass and charge parameters of the static solutions to functions of the null coordinate. The resulting matter sources, together with their conservation properties, are analyzed in detail.

In Section\,\ref{Sec5}, we derive the Vaidya-type Reissner--Nordström solution in quasitopological gravity. In Section\,\ref{Sec7}, we extend this construction to the case of a general nonlinear electrodynamics, replacing the Maxwell field with an arbitrary nonlinear electromagnetic source. In Section\,\ref{Sec7}, we specialize the analysis to quasitopological gravity coupled to Born--Infeld electrodynamics and present the corresponding Vaidya-type solutions. Finally, in Section\,\ref{Sec8}, we summarize our results and discuss possible directions for future research.

In the paper we use units in which $c=1$ and sign convention adopted in the book \cite{MTW}.

\section{QTG dilaton 2D action and field equations}\label{Sec2}

\subsection{Metric}

In what follows, we consider two-dimensional dilaton gravity derived within the framework of the QTG model. We begin by introducing the notation that will be used throughout this paper. Let $M$ be a $D$-dimensional spacetime equipped with a metric $g_{AB}$
\be \n{DDD}
ds^2=g_{AB} dX^{A} dX^B\hh A,B =0,1,\ldots, D-1 \, .
\ee
We consider the case in which the metric takes the form of the following warped product:
\be\n{GO}
ds^2=\gamma_{\mu\nu}(x) dx^{\mu} dx^{\nu}+\varphi^2(x)
d\Omega^2_\ins{D-2}\, ,
\ee
where
\be \n{OO}
d\Omega^2_\ins{D-2}=\Omega_{ij} dy^i dy^j\,
\ee
is the metric on the $(D-2)$-dimensional unit sphere $S^{D-2}$. Throughout this paper, Greek indices $(\mu,\nu,\ldots)$ take the values $0,1$, while Latin indices $(i,j,\ldots)$ run over $2,\ldots,D-1$. We denote by $\Omega_\ins{D-2}$ the volume of the unit sphere $S^{D-2}$
\be\n{OM}
\Omega_\ins{D-2}=\frac{2\pi^{(D-1)/2}}{\Gamma\big(\frac{D-1}{2}\big)} .
\ee
In particular, one has
\be
\Omega_2=4\pi \hh
\Omega_3=2\pi^2 \hh
\Omega_4=\frac{8\pi^2}{3}.
\ee

Suppose that $P_{AB}$ is a symmetric tensor respecting the spherical symmetry of the spacetime with metric~\eqref{GO}. Then it can be decomposed as
\be
P_{A}{}^{B}=
\delta^{\mu}_{A}\delta^{B}_{\nu}\,\mathcal{P}_{\mu}{}^{\nu}
+
\delta^{i}_{A}\delta^{B}_{i}\frac{\mathcal{P}}{D-2},
\ee
where $\mathcal{P}_{\mu\nu}$ is a symmetric tensor on the two-dimensional orbit space, and
$\mathcal{P}$ is a scalar.
The normalization factor $1/(D-2)$ is chosen so that $\mathcal{P}$ coincides with the trace of the angular part of the tensor, $\mathcal{P}={P}^{i}{}_{i}$.

We denote by $\nabla_A$ the $D$-dimensional covariant derivative associated with the metric $ds^{2}$, while a semicolon denotes the two-dimensional covariant derivative associated with the metric $d\gamma^{2}$. If $P_{AB}$ is conserved,
\be
\nabla_{B}P^{B}{}_{A}=0,
\ee
then its two-dimensional components satisfy
\be \n{CONS}
\frac{1}{\varphi^{D-2}}
\left(
\varphi^{D-2}\mathcal{P}_{\mu}{}^{\nu}
\right)_{;\nu}
=
\frac{1}{\varphi}\,
\varphi_{;\mu}\,\mathcal{P}.
\ee

Denote by $R_{AB}$ the Ricci tensor associated with the metric $ds^{2}$, and by
\be
\mathcal{R}_{\mu\nu}
=\frac{1}{2}\,\gamma_{\mu\nu}\,\mathcal{R}
\ee
the Ricci tensor of the two-dimensional metric $d\gamma^{2}$, where
$\mathcal{R}$ is the corresponding Ricci scalar.
For the warped-product metric \eqref{DDD} the non-vanishing
components of the $D$-dimensional Ricci tensor are
\be
\begin{split}
&R_{\mu\nu}=
\frac12\,\mathcal{R}\,\gamma_{\mu\nu}
-(D-2)\frac{\varphi_{;\mu\nu}}{\varphi}\, ,\\
&R_{ij}=\left[(D-3)(1-f)-\varphi\,\Box\varphi
\right] \Omega_{ij}\, .
\end{split}
\ee
Here
\be
\Box\varphi=\gamma^{\mu\nu}\varphi_{;\mu\nu} \,,
\qquad
f=\gamma^{\mu\nu}\varphi_{;\mu}\varphi_{;\nu}  \,.
\ee
All mixed components vanish,
\be
R_{\mu i}=0  \, .
\ee

There are only four independent scalar curvature invariants \cite{NARLIKAR} associated with a general spherically symmetrical line-element.
In the spacetime \eqref{GO} these four basic scalar invariants are constructed from the metric $\gamma_{\mu\nu}$ and a dilaton field $\varphi$\footnote{In the papers on the QTG quite often the primary curvature invariant $p$ is denoted by $\psi$. }

\be
p=\dfrac{1-f}{\varphi^2}\hhh
q=\dfrac{\Box\varphi}{\varphi}\hhh
v=\CAL{R}\hhh
u=\dfrac{\varphi^{,\mu} \varphi^{,\nu} \varphi_{;\mu\nu}}{\varphi} .
\ee
The components of the Riemann curvature tensor can be expressed in terms of these basic curvature invariants.

\subsection{Action}

The  spherically reduced action of the QTG is
\be\n{AQTG}
S_\ins{QTG}=B\CAL{S}_\ins{QTG}[\gamma,\varphi] ,
\ee
where
\ba\n{AQTG1}
\CAL{S}_\ins{QTG}&=\int d^2 x \sqrt{|\gamma|}\mathcal{L}, \\
\mathcal{L}&=\frac{1}{D-2}\varphi^{D-2} L_\ins{QTG}
\, .
\ea
Here the coefficient $B$ related to integration of angles is
\be
B=\frac{(D-2)\Omega_\ins{D-2}}{16\pi G_D}=\frac{(D-2)\Omega_\ins{D-2}}{2\kappa}\, ,
\ee
and $G_D$ is a $D$-dimensional gravitational coupling constant. The second equality defines a related parameter $\kappa$, which in four-dimensional case takes the form $\kappa=8\pi G$.

The Lagrangian density $\CAL{L}$  for the QTG model has the form (see Eq.(11) of \cite{Bueno:2025gjg})
\ba\n{LQTG}
\MC{L}=&G_2(\varphi,f)-\Box\varphi G_3(\varphi,f)+G_4(\varphi,f) \CAL{R}
\\
&-2\big(\partial_f G_4(\varphi,f)\big)\big[(\Box\varphi)^2-\varphi^{;\alpha\beta}\varphi_{;\alpha\beta}\big]\, ,
\ea
where
\be
\begin{split}
&G_2 =\varphi^{(D-2)}(D-1)h -2\varphi^{(D-2)}{p} h' \, ,  \\
&G_3=  2\varphi^{(D-3)}h'\, ,  \\
&G_4=  -\frac{1}{D-2}\varphi^{(D-2)}\lambda\, ,\\
&\lambda({p})=\frac{D-2}{2}{p}^{(D-2)/2}\int d {p}\, {p}^{-D/2} h'({p})\, .
\end{split}
\ee
Here $h=h(p)$ is a function of the primary curvature invariant $p$. The choice of this function specifies a particular quasitopological gravity (QTG) model. For the present, we leave $h(p)$ arbitrary and specify its explicit form only when required. Here and throughout the paper, a prime denotes differentiation with respect to its argument $p$.

\subsection{Field equations}

In the presence of matter, the total action takes the form
\be
S=S_\ins{QTG}+S_\ins{m} \, ,
\ee
where $S_\ins{m}$ is the action describing the matter content
\be
S_\ins{m}=\,\int d^D x \,\sqrt{-g}\,L_\ins{m}\, .
\ee
Its spherically reduced form is
\ba
&S_\ins{m}=\Omega_\ins{D-2} \CAL{S}_\ins{m}\, ,\\
&\CAL{S}_\ins{m}=\int d^2x \sqrt{\gamma}\CAL{L}_\ins{m}\hh
\CAL{L}_\ins{m}=\varphi^{D-2}L_\ins{m}\, .
\ea
So the total spherically reduced action becomes
\be
S=B\CAL{S}_\ins{QTG}+\Omega_\ins{D-2} \CAL{S}_\ins{m} .
\ee
Let us denote variations of the QTG action $\CAL{S}_\ins{QTG}$ with respect to 2D metric $\gamma_{\mu\nu}$ and the dilaton field $\varphi$ as follows
\ba\label{BGmn}
&\CAL{G}^{\mu\nu}=-\frac{2}{\sqrt{|\gamma|}}\frac{\delta \CAL{S}_\ins{QTG}}{\delta \gamma_{\mu\nu}}, \\
&\CAL{G}=-\frac{\varphi}{\sqrt{|\gamma|}}\frac{\delta \CAL{S}_\ins{QTG}}{\delta \varphi} .
\ea
We use these quantities as a left hand side of the corresponding ``modified gravity" equations. Corresponding variation of $\CAL{S}_\ins{m}$  with the negative sign we put in the right hand side of these equations and interpret them as the matter source tensor.

We define components of the reduced stress energy tensor as follows
\ba\n{SET}
\CAL{T}^{\mu\nu}&=\frac{2}{\sqrt{|\gamma|}}\frac{\delta \CAL{S}_\ins{m}}{\delta \gamma_{\mu\nu}},\\
\CAL{T}&=\frac{\varphi}{\sqrt{|\gamma|}}\frac{\delta \CAL{S}_\ins{m}}{\delta \varphi}
\, .
\ea

Note that $D$-dimensional stress-energy tensor is defined as
\be
T^{AB}=\frac{2}{\sqrt{|g|}}\frac{\delta S_\ins{m}}{\delta g_{AB}}  .
\ee
Its components in the two-dimensional sector are related to the stress-energy tensor of the dimensionally reduced theory as follows
\be
T^{\mu\nu}= \frac{1}{\varphi^{D-2}}\CAL{T}^{\mu\nu}   .
\ee

In these notations, the field equations for the QTG model coupled with matter take the form
\be \n{GEQ}
\CAL{G}^{\mu\nu}=\frac{2\kappa}{D-2}\CAL{T}^{\mu\nu}\hh \CAL{G}=\frac{2\kappa}{D-2}\CAL{T}\, .
\ee
Let us note that Eq.~\eqref{CONS}, when applied separately to the gravitational and matter sectors, takes the form
\be\label{Bianchi0}
\mathcal{G}_{\mu\alpha}{}^{;\alpha}=\frac{1}{\varphi}\varphi_{;\mu} \CAL{G}\hh
\mathcal{T}_{\mu\alpha}{}^{;\alpha}=\frac{1}{\varphi}\varphi_{;\mu} \CAL{T}\, .
\ee

For the spherically reduced two-dimensional QTG action \eqref{AQTG1} with Lagrangian density \eqref{LQTG}, a straightforward but lengthy calculation yields
\be\label{GmnCovariant}
\begin{split}
\mathcal{G}_{\mu\nu}&=-(D-1)\varphi^{D-2}\gamma_{\mu\nu}h({p})\\
&+2\varphi^{D-4}\big[-\varphi\varphi_{;\mu\nu}+\gamma_{\mu\nu}\big(1-\varphi^{;\alpha}\varphi_{;\alpha}
+\varphi \varphi^{;\alpha}_{;\alpha}
\big)\big]h'({p})\, ,\\
\CAL{G}=&-(D-1)(D-2)  \varphi^{D-2}  h({p})\\
&+ \varphi^{D-4} \big[(4D-10)\big(1-\varphi^{;\alpha}\varphi_{;\alpha}\big)+2(D-3)\varphi\varphi_{;\alpha}^{;\alpha}\\
&-\CAL{R}\varphi^2
\big] h'({p})
+\varphi^{D-6}\big[-2\varphi^2\varphi^{;\beta}_{;\beta}\varphi_{;\alpha}^{;\alpha}+2\varphi^2\varphi^{;\beta \alpha}\varphi_{;\beta \alpha}\\
&-4\varphi^{;\beta}\varphi_{;\beta}\varphi^{;\alpha}\varphi_{;\alpha}+8\varphi^{;\beta}\varphi_{;\beta}
-4\\
&+4 \varphi\varphi_{;\beta}^{;\beta}\varphi^{;\alpha}\varphi_{;\alpha}
-4\varphi \varphi_{;\beta}^{;\beta}
\big] h''({p})\, .
\end{split}
\ee


\section{Generalized Birkhoff theorem}\label{Sec3}

It has been shown that the Birkhoff theorem holds for a broad class of vacuum quasitopological gravity (QTG) models \cite{Bueno:2025qjk}. In this paper, we extend this result to quasitopological gravities coupled to matter whose stress--energy tensor satisfies a specific set of conditions. In particular, we demonstrate that the Birkhoff theorem remains valid for QTG interacting with both linear and nonlinear electrodynamics.

Suppose the two-dimensional part of matter stress-energy tensor satisfies a condition
\be \n{TTTT}
\CAL{T}_{\mu\nu}=\dfrac{1}{2}\CAL{T}^{\alpha}_{\alpha}\gamma_{\mu\nu}\, .
\ee
Let us show that the spherically symmetric gravitational field generated by this matter distribution, as determined by the QTG field equations, possesses a Killing vector. To establish this result, we first note that the tensor $\CAL{G}_{\mu\nu}$ defined in \eqref{GmnCovariant} has the following structure:
\be
\CAL{G}_{\mu\nu}=\CAL{A}\varphi_{;\mu\nu}+\CAL{B} \gamma_{\mu\nu}\, ,
\ee
Thus for the matter stress-energy tensor \eqref{TTTT} the field equation \eqref{GEQ} imply that
\be
\varphi_{;\mu\nu}=\CAL{W}\gamma_{\mu\nu},
\ee
where $\CAL{W}$ is a scalar function.

Define the two dimensional Levi-Civita tensors
\be
e_{\mu\nu}=\sqrt{-\gamma}\epsilon_{\mu\nu},\hskip 1cm e^{\mu\nu}=-\frac{1}{\sqrt{-\gamma}}\epsilon^{\mu\nu} .
\ee
Then the vector
\be\n{KIL}
\xi^\mu=e^{\mu\epsilon}\varphi_{;\epsilon}
\ee
has the following property
\be
\xi_{\mu;\nu}=e_{\mu}{}^\epsilon\varphi_{;\epsilon\nu}=e_{\mu}{}^\epsilon\gamma_{\epsilon\nu} \CAL{W} =e_{\mu\nu}\CAL{W}  .
\ee
Hence  the vector $\xi^{\mu}$ satisfies the two dimensional Killing equation
\be\n{xi}
\xi_{\mu;\nu}+\xi_{\nu;\mu}=0  .
\ee
One also has
\be \n{xv}
\xi^{\mu}\varphi_{,\mu}=e^{\mu\nu}\varphi_{,\mu}\varphi_{,\nu}=0\, .
\ee
These two relations \eqref{xi} and \eqref{xv} imply that the vector $\xi^{A}=\delta^A_{\mu}\xi^{\mu}$ is the Killing vector of the spacetime with metric \eqref{GO}.

Consider integral lines $x^{A}=x^{A}(t)$ of this Killing vector
\be
\dfrac{dx^{A}}{dt}=\xi^A\, .
\ee
In the domain where $\xi^{A}$ is either timelike or spacelike using the Killing parameter $t$ one can write the metric \eqref{GO} in the following standard form
\be\n{STAT}
ds^2=-N(r)^2 f(r) dt^2+\frac{1}{f(r)}dr^2+r^2 d\Omega^2_\ins{D-2} \,,
\ee
where $\varphi=r$.
Let us use the definition \eqref{KIL} and calculate the square of the Killing vector $\xi^{\mu}$
\be\n{NNNN}
\xi^{\alpha}\xi_{\alpha}=e^{\alpha\beta}e_{\alpha\epsilon}\varphi_{;\beta}\varphi^{;\epsilon}
=-\delta^{\beta}_{\epsilon}\varphi_{;\beta}\varphi^{;\epsilon}=-f \, .
\ee
Since $\xi^{\alpha}\xi_{\alpha}=-N^2 f$, relation \eqref{NNNN} shows that $N=1$.

Let us summarize. Any spherically symmetric solution of the QTG equations for the matter distribution satisfying relation \eqref{TTTT} possesses a Killing vector, and being written in the form \eqref{STAT} it has metric coefficient $N=1$.


\section{Vaidya-type solutions of QTG equations}\label{Sec4}

\subsection{Equations}

Let us now consider a special class of interesting time-dependent solutions of the QTG equations. To this end, we first rewrite the static metric \eqref{STAT} in advanced-time coordinates. Specifically, we introduce\footnote{
Instead of incoming null coordinate $v$ one can consider a retarded time coordinate $u$
\be\nonumber
du=dt-\dfrac{dr}{fN}\, .
\ee
All the results obtained in the section and later can be easily written in $(u,r)$ coordinates.
}
\be
dv=dt+\dfrac{dr}{fN}\, .
\ee
In coordinates $(v,r)$ the metric \eqref{STAT} takes the form
\be \n{NFR}
ds^2=-N^2 f dv^2+2 N dv dr +r^2 d\Omega^2_\ins{D-2}\, .
\ee
We now consider the metric in the above form, allowing the functions $f$ and $N$ to depend on both $v$ and $r$, i.e., $f=f(v,r)$ and $N=N(v,r)$. Our goal is to find solutions of the QTG field equations within this class of metrics.

For  arbitrary spacetime dimension $D$ the  two-dimensional tensor part $\mathcal{G}^{\mu\nu}$  in  $v$-$r$ coordinates reads
\ba\n{RV_SET}
\mathcal{G}_{vv}&=(D-1)r^{D-2} fN^2 h({p})\\
&+r^{D-4}N\big[
-r N f f_{,r}-2(1-f) f N -r f_{,v}\big] h'({p})\, ,\\
\mathcal{G}_{vr}&=-(D-1) r^{D-2}N h({p})\\
&+r^{D-4} N\big[
r f_{,r}+2(1-f)\big]  h'({p}) \, ,\\
\mathcal{G}_{rr}&=2r^{D-3}\frac{N_{,r}}{N} h'({p}) \, .
\ea
We denote
\be
H=r^{D-1}h\, .
\ee
Using this definition together with appropriate combinations of the relations in \eqref{RV_SET}, this system of equations can be simplified considerably, yielding
\be \n{EQQ}
\begin{split}
&\mathcal{G}_{vv}+N f \mathcal{G}_{vr}=N H_{,v}\, ,\\
&\mathcal{G}_{vr}=-N H_{,r}\, ,\\
&\mathcal{G}_{rr}=2r^{D-3}\frac{N_{,r}}{N} h'({p})\, .
\end{split}
\ee

\subsection{Matter ansatz}

In what follows we consider the following ansatz for the two-dimensional part of the stress energy tensor $T_{AB}$ of the matter is of the form
\be \n{SETFLUX}
T_{\mu\nu}=\frac{1}{r^{D-2}}\CAL{T}_{\mu\nu}   \hh
\CAL{T}_{\mu\nu}=T\gamma_{\mu\nu}+\sigma v_{,\mu} v_{,\nu}\, .
\ee
For the angular part of the stress-energy tensor, we retain the form
$\mathcal{T}^{i}_{j}=\dfrac{1}{D-2}\delta^{i}_{j},\mathcal{T}$.
We now assume that the functions $T$, $\mathcal{T}$, and $\sigma$ depend on both coordinates, $v$ and $r$.
For a stress-energy tensor of this form, the conservation law \eqref{CONS} implies the following relations:
\be \n{EEQQ}
\begin{split}
&\sigma_{,r}=-NT_{,v}\, ,\\
&
T_{,r}=r^{-1}\CAL{T}\, .
\end{split}\ee

For the stress-energy tensor \eqref{SETFLUX} one has
\be
\begin{split}
&\mathcal{T}_{vv}+N f \mathcal{T}_{vr}=\sigma\, ,\\
&\mathcal{T}_{vr}=N T\, ,\\
&\mathcal{T}_{rr}=0\, .
\end{split}
\ee
Since $\mathcal{T}_{rr}=0$, the $rr$ component of the field equations \eqref{EQQ} implies that
$N_{,r}=0$. Then we can always choose the gauge $N=1$.
Substituting this result into the remaining field equations, we obtain
\be \n{HHVR}
H_{,v}=\frac{2\kappa}{D-2}\sigma \,\hh
H_{,r}=-\frac{2\kappa}{D-2}T\, .
\ee
Equations \eqref{EEQQ} with $N=1$ play the role of the integrability conditions for the equations \eqref{HHVR}.

Let us note that in the absence of fluxes, i.e., when $\sigma=0$, both $H$ and $h$ are independent of time. The resulting spacetime is therefore static and satisfies the generalized Birkhoff theorem.


\section{Vaidya-Type Reissner--Nordström Solution in QTG}\label{Sec5}

As a first example we consider a Vaidya-type Reissner--Nordström solution in QTG. In the absence of null flux this solution is nothing but a  QTG analogue of a higher dimensional Reissner--Nordström metric. This metric and its properties were discussed in \cite{PinedoSoto:2026hfm}, where further references can be found (see also \cite{hao2026chargedregularblackholes}).

The linear Maxwell action in $D$ dimensions is
\begin{equation}
\begin{split}
 &S_\ins{m}=-\frac{1}{16\pi}\int d^D X\sqrt{-g}\,\CAL{F}\, ,\\
 &F_{AB}=2\nabla_{[A}A_{B]} \hh
 \CAL{F}=F_{AB}F^{AB} \, .
 \label{Maxwell_action}
 \end{split}
\end{equation}
If an external charged current is present, the interaction term is
\begin{equation}
 S_{\rm int}=-\int d^D X\sqrt{-g}\,A_A j^A \, .
 \label{interaction_action}
\end{equation}
Variation with respect to $A_A$ gives
\begin{equation}
 \nabla_B F^{AB}=4\pi j^A\, ,
 \qquad
 \nabla_{[A}F_{BC]}=0 \, .
 \label{Maxwell_equations}
\end{equation}
The Maxwell stress-energy tensor is
\begin{equation}
T_{AB} =\frac{1}{4\pi}\left(
 F_{AC}F_B{}^C-\frac{1}{4}g_{AB}\CAL{F}
 \right).
 \label{Maxwell_stress}
\end{equation}
It obeys
\begin{equation}
 \nabla_B T_A{}^{B}=-F_{AB}j^B .
 \label{Maxwell_nonconservation}
\end{equation}

For a spherically symmetric solution in the metric \eqref{NFR} the only nonvanishing component of the field is
\be
F_{rv}=-F_{vr}=E\, .
\ee
The $B=v$ component of the Maxwell equation (\ref{Maxwell_equations}) outside a source, i.e. with the current $j^v=0$, is
\begin{equation}
 \partial_r\Big(\frac{1}{N}r^{D-2}E\Big)=0
\end{equation}
and hence its solution
\begin{equation}\label{E_solution}
 E(v,r)=N \frac{Q(v)}{r^{D-2}} \, .
\end{equation}
The $B=r$ component of the Maxwell equation gives
\begin{equation}
\partial_v\Big(N r^{D-2}  F^{rv}   \Big)=4\pi N r^{D-2} j^r .
\end{equation}
Using $F^{rv}=-E/N^2$ and \eqref{E_solution}, one obtains
\begin{equation}
 j^v=0\, ,
\hskip 0.5cm
 j^r  =-\frac{\dot Q(v)}{4\pi N r^{D-2}}\, ,
\hskip 0.5cm
 j^i=0 \, .
 \label{current}
\end{equation}
where $\dot Q\equiv\partial_v Q$ . Thus, for $\dot Q>0$, the charged current is directed along $-\partial_r$, as
appropriate for an ingoing charged flux in the coordinates $(v,r)$.
The current is conserved:
\begin{equation}
 \nabla_A j^A=\frac{1}{N r^{D-2}}\partial_r\left( N r^{D-2} j^r\right)=0 \, .
\end{equation}

The stress-energy tensor of the electric field gives for  quantities $T$ and $\CAL{T}$ the following expressions
\be
T=-\dfrac{E^2}{8 \pi N^2} r^{D-2}\hh
\CAL{T}=\dfrac{E^2}{8 \pi N^2}r^{D-2}\, .
\ee
The $rr$ component of the stress--energy tensor vanishes. As a result, the gravitational field equations imply that $N'=0$. Therefore, without loss of generality, we may choose the gauge in which $N=1$. This gauge choice will be assumed throughout the remainder of this subsection.
If the charge $Q$ depends on $v$, then conservation of the total energy-momentum tensor implies
\begin{equation}\n{SSSS}
 \sigma(v,r)=\chi(v)
 -\frac{Q(v)\dot Q(v)}{4\pi(D-3)r^{D-3}} .
\end{equation}
Here $\chi(v)$ is an arbitrary function describing a flux of neutral null radiation. Integrating the first equation in \eqref{HHVR}, one obtains
\be \n{HHhh}
H=\mu(v)-\frac{\kappa}{4\pi(D-2)(D-3)}\frac{Q^2(v)}{r^{D-3}}\, ,
\ee
where $\mu_{,v}$ is proportional to $\chi(v)$. The function $\mu$
is related to the physical mass $M(v)$ measured at infinity as follows
\be
\mu(v)=\frac{2\kappa M(v)}{(D-2)\Omega_\ins{D-2}} \, .
\ee

Differentiating this expression with respect to $r$ and substituting the result into the second equation in \eqref{HHVR},
we find that the latter is identically satisfied.
Using \eqref{HHhh} one finds
\be\n{hhhhh}
h=\frac{\mu(v)}{r^{D-1}}-\frac{\kappa}{4\pi(D-2)(D-3)}\frac{Q^2(v)}{r^{2D-4}}\, .
\ee
Once the function $h(v,r)$ has been determined, the metric can be reconstructed as follows. Inverting the relation $h=h(p)$ yields $p=p(h)$. Substitution of \eqref{hhhhh} into this inverse relation then gives $p=p(v,r)$. The metric function $f$ is subsequently obtained from
\be
f=1-r^2 p(v,r)\, .
\ee


\section{Vaidya-Type Solutions in QTG Coupled to Nonlinear Electrodynamics}\label{Sec6}

\subsection{Action and field equations}

Let us demonstrate that the approach developed in the previous section can be readily extended to construct Vaidya-type solutions in quasitopological gravity (QTG) coupled to nonlinear electrodynamics. Static spherically symmetric solutions of this theory were obtained and analyzed in \cite{PinedoSoto:2026hfm}. We now show that the formalism can be naturally generalized to include charge-carrying null currents and null fluid fluxes, thereby yielding a broad class of dynamical charged solutions.

Let consider a non-linear electrodynamics
\footnote{
For useful references on this subject see e.g.
\cite{BornInfeld, Ketov:2001dq, Kerner:2001qq, Sorokin:2021tge, Yang:2023BI, SingularitiesNED}.
}
whose action is
\begin{equation}
 S_\ins{m}=-\frac{1}{16\pi}\int d^D X\sqrt{-g}\,{L}(\CAL{F}) \, .
 \label{NED_action}
\end{equation}
Here as earlier    $\CAL{F}=F_{AB}F^{AB}$.
The Maxwell theory is recovered for
\begin{equation}\label{L_BI}
{L}(\CAL{F})=\CAL{F}\, .
\end{equation}
If an external charged current is present, the interaction term is
\begin{equation}
S_{\rm int}=-\int d^D X\sqrt{-g}\,A_Aj^A .
 \label{interaction_action1}
\end{equation}
We denote
\begin{equation}
K(\CAL{F})=\frac{d {L}}{d \CAL{F}} \, .
 \label{P_definition}
\end{equation}
Variation with respect to $A_A$ gives
\begin{equation}\label{NED_equation1}
\nabla_B\left(K F^{AB}\right)=4\pi j^A\, .
\ee
As usual, the second set of equations,
\be
\nabla_{[A}F_{BC]}=0\, ,
\ee
guarantees the existence of a vector potential $A_A$ such that
$F_{AB}=2\nabla_{[A}A_{B]}$.
It is also convenient to introduce the electric displacement tensor
\begin{equation}\n{DDDD}
 D^{AB}=K F^{AB} .
\end{equation}
Then Eq.\eqref{NED_equation1} takes the Maxwell form
\begin{equation}\label{DAB}
 \nabla_B D^{AB}=4\pi j^A .
\end{equation}
The stress-energy tensor is
\begin{equation}
 T_{AB}^\ins{NED}=\frac{1}{4\pi}\left(
 K F_{AC}F_B{}^C-\frac{1}{4}g_{AB}{L}
 \right) .
 \label{NED_stress}
\end{equation}

\subsection{Spherically symmetric solutions}

For a spherically symmetric electric field the tensor of the field strength has only one non-vanishing component
\be
F_{rv}=-F_{vr}=E(v,r)
\ee
and one has $\CAL{F}=-2E^2/N^2$.
The corresponding displacement tensor $D_{AB}$ also has only one non-vanishing component $D_{rv}$ which we denote by $D$, and \eqref{DDDD} implies
\be
D=K E\, .
\ee

The $B=v$ field equation with the current $j^v=0$  gives
\begin{equation}
 \partial_r\Big(\frac{1}{N}r^{D-2} K E\Big)=0 \, .
\end{equation}
Thus
\begin{equation}
D\equiv  K E=N\frac{Q(v)}{r^{D-2}} \, .
 \label{E_implicit}
\end{equation}
This is, in general, an algebraic equation for $E$.
Its solution can be written in the form
\be
E=E(r,Q)\, ,
\ee
which is valid in both cases: when $Q=Q(v)$ and $Q=$const.

The $B=r$ field equation gives
\begin{equation}
 \partial_v\left(N r^{D-2}K F^{rv}\right)=4\pi N r^{D-2} j^r .
\end{equation}
Since $F^{rv}=-E/N^2$  and $r^{D-2} K E= N Q(v)$, one finds for $r>0$
\begin{equation}
 j^v=0\, ,
 \qquad
 j^r=-\frac{\dot Q(v)}{4\pi N r^{D-2}}\, ,
 \qquad
 j^i=0\, .
 \label{current1}
\end{equation}
Therefore the current is conserved:
\begin{equation}
 \nabla_Aj^A=\frac{1}{N r^{D-2}}\partial_r\left(N r^{D-2} j^r\right)=0 \, .
 \label{current_conservation}
\end{equation}
A convenient gauge potential is
\begin{equation}
 A_A d X^A=-\Phi(r,Q(v))d v\, ,
 \qquad
 E=-\partial_r\Phi \, .
 \label{potential_definition}
\end{equation}
The electric potential $\Phi$ vanishing at the infinity can be written as follows
\begin{equation}
 \Phi=\int_r^\infty E(r',Q) d r' .
 \label{NED_potential}
\end{equation}
For a general non-linear theory this potential is determined by solving
\eqref{E_implicit} for $E$ and then substituting the result into
\eqref{NED_potential}.

Let us summarize. In both Maxwell and nonlinear electrodynamics, Gauss's law remains valid. Namely, the flux of the electric displacement field through any closed $(D-2)$-dimensional surface enclosing the charge is independent of the choice of the surface and remains constant in time until a charge crosses it. For spherically symmetric configurations, this property uniquely determines the field strength in terms of the enclosed charge (see \eqref{E_solution} and \eqref{E_implicit}). Consequently, for a fixed charge, the corresponding solutions are necessarily static. This result is the electrodynamic analogue of the gravitational Birkhoff theorem.

The conserved stress-energy tensor in the presence of fluxes satisfies the condition $T_{rr}=0$, which allows one to choose the gauge $N=1$. The stress-energy tensor of the nonlinear electromagnetic field is of the form \eqref{SETFLUX}, and its nonvanishing components are
\ba\label{TcaTsigma}
&T=-\frac{1}{4\pi}\left(K E^2+\frac{{L}}{4}\right)r^{D-2},\\
&\CAL{T}=-\frac{{L}}{16\pi}r^{D-2}\, ,\\
&\sigma=\chi(v)-\frac{\dot{Q}(v)\Phi(r,Q)}{4\pi} .
\ea
Integrating the first equation in \eqref{HHVR} one gets
\be \n{HInfeld}
H=\mu(v)-\frac{2\kappa}{D-2}\frac{1}{4\pi}\int^v dv\,\dot{Q}(v)\Phi(r,Q)\, .
\ee
The second equation in \eqref{HHVR} allows one to fix an arbitrary function of $r$ in H.

Let us note that the integration over $v$ in this expression can be
replaced by an integration over the electric charge. Indeed, after
determining $\Phi(r,Q)$, the function $H$ can be obtained by integrating
with respect to charge from $0$ to its final value $Q$ while keeping the
radius $r$ fixed.

It is also worth noting that the second relation in \eqref{HHVR}
provides an alternative way of determining $H$. Namely, for a fixed
value of the charge $Q$, one can obtain $H$ by integrating the function
$T$ with respect to the radial coordinate $r$. This representation
remains valid when the charge becomes time dependent, $Q=Q(v)$.

The above result can be summarized in the following statement. To
construct a Vaidya-type solution of the quasitopological gravity
equations coupled to nonlinear electrodynamics, it is sufficient to
first obtain the corresponding static solution and then promote the
constant electric charge to an arbitrary function of the advanced time,
$Q \rightarrow Q(v)$.

Using this result one obtains the following expression for $h(p)$
\be \n{hBorn}
h=\frac{1}{r^{D-1}}\Big[\mu(v)-\frac{2\kappa}{(D-2)}\frac{1}{4\pi}\int^v dv\,\dot{Q}(v)\Phi(r,Q)\Big]\, .
\ee
To reconstruct the metric function $f$ corresponding to the obtained solution, one follows the procedure described at the end of the previous section.

\subsection{Remark on the cosmological constant}

In the previous calculations we assumed that the solutions we are dealing with are asymptotically flat. Let us show that all the obtained results can be easily generalized to a case when the spacetime has either deSitter or anti-deSitter asymptotics.

For this purpose we add the cosmological constant term to the gravity action (\ref{QTA}).
\ba \n{SL}
S_\Lambda&=-\frac{1}{\kappa}\int d^D x
\sqrt{|g|}\Lambda
\ea
After the spherical reduction $S_\Lambda=B\CAL{S}_\Lambda$ the corresponding two-dimensional action
$\CAL{S}_\Lambda$ becomes
\ba \n{CALSL}
\CAL{S}_\Lambda&=-\frac{2}{D-2}\int d^2 x
\sqrt{|\gamma|}\varphi^{D-2}\Lambda \, .
\ea

The cosmological constant term may be incorporated either into the left-hand side of the gravitational field equations or transferred to the right-hand side, where it can be regarded as a special form of matter. For the purposes of the present analysis, the latter formulation is more convenient.
In this representation, the action \eqref{CALSL} generates a diagonal stress--energy tensor with the following two-dimensional part:
\be
\MC{T}_{\Lambda}{}_{\mu\nu}=-\frac{2}{D-2}\gamma_{\mu\nu}\varphi^{D-2}\Lambda \, .
\ee
Therefore
\ba
T_{\Lambda}&=-\frac{2}{D-2}\varphi^{D-2}\Lambda\, ,
\\
\MC{T}_{\Lambda}&=-2\varphi^{D-2}\Lambda \, .
\ea
Taking into account the normalization factors in the definitions of the actions $\CAL{S}_\ins{QTG},$ $\CAL{S}_\ins{m}$, and $\CAL{S}_\Lambda$ we obtain the following gravity equations
\ba
\CAL{G}_{\mu\nu}&=\CAL{T}_{\Lambda}{}_{\mu\nu}+\frac{2\kappa}{D-2}\CAL{T}_{\mu\nu}\, ,\\
\CAL{G}&=\CAL{T}_{\Lambda}+\frac{2\kappa}{D-2}\CAL{T}\, .
\ea

Adding the cosmological constant results in the following shift of the components of the stress-energy tensor \eqref{EEQQ}
\be\nonumber
T\to T+ \frac{D-2}{2\kappa} T_{\Lambda}\hh \CAL{T}\to \CAL{T}+\frac{D-2}{2\kappa}\CAL{T}_{\Lambda}\hh \sigma\to \sigma\, .
\ee
In the absence of fluxes, that is, when $\sigma=0$, this modification preserves the conditions required by Birkhoff's theorem. Consequently, the metric remains static.

It is easy to check that in the presence of the cosmological constant one gets
\be
h(p)\to h(p)-\frac{2}{(D-2)(D-1)}\Lambda \, .
\ee


\section{An example}\label{Sec7}

To illustrate a developed approach let us consider special interesting Vaidya-type solutions.

First, let us note that a particular model of quasitopological gravity (QTG) coupled to nonlinear electrodynamics is completely specified by the choice of two functions,
\be
h=h(p)  \hh
{D}={D}(E)\, .
\ee
The first function, $h=h(p)$, characterizes the QTG sector of the theory, whereas the second, the electric displacement function ${D}={D}(E)$, specifies the nonlinear electrodynamics model.

For regular vacuum black holes in QTG two choices of $h(p)$ are often considered
\begin{itemize}
\item Hayward type QTG: $h(p)=\dfrac{p}{1+\ell^2 p}$\, .
\item Born-Infeld type QTG: $h(p)=\dfrac{p}{\sqrt{1+\ell^4 p^2}}$.
\end{itemize}

For the first choice of the function $h(p)$, the corresponding vacuum QTG solution reproduces the Hayward metric ansatz for a regular black hole.
However, it was shown in \cite{PinedoSoto:2026hfm} that, for charged black hole solutions in QTG, this choice does not, in general, guarantee regularity of the solution. In particular, for a wide range of values of the charge parameter, the principal curvature invariant $p$ becomes negative at sufficiently small values of $r$. As $r$ decreases further, $p$ continues to decrease until it reaches the critical value $p=-1/\ell^2$, where a curvature singularity develops inside the black hole.

For this reason, to illustrate Vaidya-type solutions in QTG, we shall consider the Born--Infeld model, for which such interior singularities are absent\footnote{
Interestingly, in these models the standard de Sitter--like core is replaced by an anti-de Sitter--like core \cite{PinedoSoto:2026hfm}.}
.

\subsection{Born-Infeld electrodynamics}

As for the choice of a non-linear electrodynamics we use the Born-Infeld model. This model is widely known and its properties are described in very details in many books and reviews (see e.g \cite{Ketov:2001dq, Kerner:2001qq, Sorokin:2021tge, Yang:2023BI}). For this model the Lagrange density $L$  (\ref{L_BI}) is of the form
\be
L=4 b^2\Big(\sqrt{1+\frac{1}{2b^2}\CAL{F}}-1\Big)\, .
\ee
For a spherically symmetric solution in the gauge $N=1$ the electric displacement $D$, electric field $E$, and the field invariant $\CAL{F}$ are
\be
\begin{split}
&D=\frac{E}{\sqrt{1-\frac{E^2}{b^2}}}= \frac{Q(v)}{r^{D-2}}\, , \\
&{E}=\frac{Q(v)}{\sqrt{r^{2D-4}+\frac{Q^2(v)}{b^2}}} \, ,\\
&\CAL{F}=-2\frac{Q^2(v)}{r^{2D-4}+\frac{Q^2(v)}{b^2}}\, .
\end{split}
\ee

The electric potential
\ba
\Phi(r,Q)=\int_r^{\infty} d r' E(r',Q) \, ,
\ea
can be written in terms of hypergeometric function ${}_2F_{1}$
\ba
\Phi&=\frac{Q(v)\, {}_{2}F_{1}\Big( \frac{1}{2}, \frac{1}{2}-\beta; \frac{3}{2}-\beta;-z\Big)}{(D-3)r^{D-3}} \, .
\ea
Here we used the notations
\be
\beta=\frac{1}{2(D-2)} \hh z=\frac{Q^2(v)}{b^2 r^{2D-4}}  \, .
\ee
At large radii the potential falls down as follows
\be
\Phi(r,Q)\big|_{r\to\infty}\to \frac{Q(v)}{(D-3)r^{D-3}}+O(r^{-3D+7})\, .
\ee

The stress-energy tensor (\ref{TcaTsigma}) for this model reads
\ba
T&=\frac{b^2 r^{D-2}}{4\pi}\big(1-\sqrt{1+z}\big)\, ,
\\
\CAL{T}&=\frac{b^2 r^{D-2}}{4\pi}\big(1-\frac{1}{\sqrt{1+z}}\big)\, ,
\\
\sigma&=\chi(v)-\frac{\dot{Q}(v)\Phi(r,Q(v))}{4\pi}\, .
\ea
It can be explicitly checked that the conditions $\sigma_{,r}=-T_{,v}$
and $T_{,r}=r^{-1}\CAL{T}$ are satisfied.

The integral over $v$ entering the EQ. (\ref{hBorn}) can be computed explicitly
\ba
\int^v & dv\,\dot{Q}(v)\Phi(r,Q)=\int_0^{Q(v)} dQ Q\Phi(r,Q)\\
=&\frac{b^2 r^{D-1}\Big[1-{}_{2}F_{1}\Big( -\frac{1}{2}, -\frac{1}{2}-\beta; \frac{1}{2}-\beta;-z\Big)\Big]}{D-1} ,
\ea
where the lower integration limit $Q=0$  was chosen such that the integral decreases at large radii.

Thus the solution for the function $h$ (\ref{hBorn}) becomes
\ba \n{hBorn1}
h&=\frac{\mu(v)}{r^{D-1}}\\
&-\frac{\kappa b^2 \Big[1-{}_{2}F_{1}\big( -\frac{1}{2}, -\frac{1}{2}-\beta; \frac{1}{2}-\beta;-z\big)\Big]}{2\pi(D-2)(D-1)}\, .
\ea

One can check that function $H=r^{D-1} h$ satisfies the second relation in Eq. (\ref{HHVR}), what is the consequence of the integrability conditions (\ref{EEQQ}).
In the static case, when $Q(v)=Q$ and $\mu(v)=\mu$, the above solution reduces to that of \cite{PinedoSoto:2026hfm} upon identifying the charge parameter used in the present paper with $\sqrt{4\pi} Q$ in the notations of \cite{PinedoSoto:2026hfm}\footnote{
Note that in the present paper we use Gaussian units, whereas \cite{PinedoSoto:2026hfm} employs Heaviside--Lorentz units. This difference in conventions accounts for the additional factor of $4\pi$.
}.

In the limit $b\to\infty$ non-linear electrodynamics reduces to the Maxwell theory. In this limit
the parameter $z\to 0$ and, hence,
\be
\Phi\big|_{b\to \infty}=\frac{Q(v)}{(D-3)r^{D-3}}+\ldots \, .
\ee

The generalization to the QTG theory with the cosmological constant boils down to the following form of
\ba \n{hBornLambda}
h&=\frac{\mu(v)}{r^{D-1}}-  \frac{2}{(D-2)(D-1)}\Lambda     \\
&-\frac{\kappa b^2 \Big[1-{}_{2}F_{1}\big( -\frac{1}{2}, -\frac{1}{2}-\beta; \frac{1}{2}-\beta;-z\big)\Big]}{2\pi(D-2)(D-1)}\, .
\ea

In the limit $b\to \infty$ (Maxwell field)
\ba \n{hBorn2}
h\big|_{b\to \infty}&=\frac{\mu(v)}{ r^{D-1}}-  \frac{2\kappa}{(D-2)(D-1)}\Lambda   \\
&-\frac{\kappa}{4\pi(D-2)(D-3)}\frac{Q^2(v)}{r^{2D-4}}+\ldots\, .
\ea

\subsection{"Double Born-Infeld model"}\label{Subsection71}

Let us apply the obtained result to the Born-Infeld type of QTG with the cosmological constant and the matter described by the Born-Infeld electrodynamics.
We call such a theory ``double Born-Infeld model".
In the model
the function $h$ is given by (\ref{hBornLambda}). The Born-Infeld quasitopological gravity is defined by the function
\be
h(p)=\dfrac{p}{\sqrt{1+\ell^4 p^2}}\, ,
\ee
where $\ell$ is the fundamental length scale of the theory. Let us express the basic curvature invariant $p$ in terms of $h$
\be
p=\frac{h}{\sqrt{1-\ell^4 h^2}}\, .
\ee
The metric is given by the solutions $N=1$ and $f=1-r^2 p$. Thus we arrive to the explicit expression
\be
f=1-\frac{r^2 h}{\sqrt{1-\ell^4 h^2}}\, ,
\ee
where the function $h$ is given by (\ref{hBornLambda}).


\section{Discussion}\label{Sec8}

In this paper we have developed a general method for constructing
Vaidya-type solutions in quasitopological gravity coupled to nonlinear
electrodynamics in arbitrary spacetime dimensions.
Our construction starts from the corresponding static, spherically
symmetric charged solutions and is based on a remarkably simple
observation. After rewriting the static metric in advanced (or
retarded) null coordinates, the integration constants characterizing
the solution can be promoted to arbitrary functions of the corresponding
null coordinate. The resulting time dependence is consistently supported
by physically natural matter sources describing a null energy flux
together with a charge-carrying null current.\footnote{
We have established this result for quasitopological gravity coupled to
nonlinear electrodynamics. Nevertheless, the underlying construction is
expected to have a considerably broader range of applicability. One can expect that whenever
the matter sector has a stress-energy tensor of the form
\eqref{SETFLUX} and the corresponding static solution is characterized
by integration constants, a Vaidya-type generalization in advanced
coordinates can be obtained simply by promoting these integration
constants to arbitrary functions of the advanced time $v$.}
We first derived the field equations for a general class of
spherically symmetric spacetimes in the framework of the
two-dimensional dilaton formulation of quasitopological gravity.
Using these equations, we established a generalized version of
Birkhoff's theorem showing that, in the absence of null fluxes if the two=dimensional part of the stress-energy tensor has the form \eqref{TTTT}
the solutions are necessarily static. We then considered the
time-dependent metric ansatz and demonstrated that the field
equations simplify considerably when written in terms of the function
$H=r^{D-1}h$, reducing the problem to a pair of first-order equations
that directly are determined by the required matter sources.

The formalism was illustrated first for linear Maxwell
electrodynamics coupled to QTG, where it reproduces the natural higher-dimensional
generalization of the charged Vaidya solution. We then extended the
construction to a general nonlinear electromagnetic theory.
The varying electric charge is accompanied by a conserved null
electric current, while conservation of the total stress-energy tensor
uniquely determines the additional null energy flux required to
support the dynamical geometry. Thus every static charged solution of
QTG coupled to nonlinear electrodynamics immediately generates a
corresponding radiating solution.

As an application, we discussed models based on Born--Infeld-type
quasitopological gravity. Unlike the Hayward-type models, whose
charged solutions generally develop interior curvature singularities, the Born--Infeld QTG models remain regular
throughout the spacetime. They therefore provide a particularly
natural setting for studying dynamical regular black holes with
time-dependent mass and charge.

The solutions obtained in this paper provide a useful framework for
investigating a variety of dynamical processes in higher-curvature
gravity. Possible applications include gravitational collapse,
accretion onto regular black holes, evaporation driven by null
radiation, and the evolution of charged compact objects in modified
gravity theories. This approach can
be applied to a broad class of theories and should also admit further
generalizations, for example to asymptotically (anti-)de Sitter
spacetimes, different horizon topologies, and more general matter
sources.

An interesting direction for future work is the investigation of the
global properties and causal structure of these dynamical solutions,
including the evolution of trapping horizons, thermodynamic
properties, and the stability of radiating regular black holes in
quasitopological gravity.


\acknowledgments

This work was supported by the Natural Sciences and Engineering Research Council of Canada.
The authors (V.F. and A.Z.) are also grateful to the Killam Trust for its financial support.
This work was supported by JSPS KAKENHI Grant Numbers JP25K07281 (C.Y.) and JP24K07027 (C.Y.).



%


\end{document}